\documentclass[aps,prl,twocolumn,showpacs,superscriptaddress,groupedaddress]{revtex4}  
\usepackage{graphicx}  
\usepackage{dcolumn}   
\usepackage{bm}        
\usepackage{amssymb}   
\usepackage{amsmath}
\begin{document}

\title{Magnetic field control of cycloidal domains and electric polarization in multiferroic BiFeO$_3$}

\author{S. Bord\'acs}
\affiliation{Department of Physics, Budapest University of Technology and Economics and MTA-BME Lend\"ulet Magneto-optical Spectroscopy Research Group, 1111 Budapest, Hungary}
\affiliation{Hungarian Academy of Sciences, Premium Postdoctor Program, 1051 Budapest, Hungary}
\author{D. G. Farkas}
\affiliation{Department of Physics, Budapest University of Technology and Economics and MTA-BME Lend\"ulet Magneto-optical Spectroscopy Research Group, 1111 Budapest, Hungary}
\author{J. S. White}
\affiliation{Laboratory for Neutron Scattering and Imaging (LNS), Paul Scherrer Institut (PSI), CH-5232 Villigen, Switzerland}
\author{R. Cubitt}
\affiliation{Institut Laue-Langevin, 71 avenue des Martyrs, CS 20156, 38042 Grenoble cedex 9, France}
\author{L. DeBeer-Schmitt}
\affiliation{Oak Ridge National Laboratory, Oak Ridge, Tennessee 37831, USA}
\author{T. Ito}
\affiliation{National Institute of Advanced Industrial Science and Technology (AIST), Tsukuba, 305-8562 Ibaraki, Japan}
\author{I. K\'ezsm\'arki}
\affiliation{Department of Physics, Budapest University of Technology and Economics and MTA-BME Lend\"ulet Magneto-optical Spectroscopy Research Group, 1111 Budapest, Hungary}
\affiliation{Experimental Physics V, Center for Electronic Correlations and Magnetism, University of Augsburg, 86159 Augsburg, Germany}

\date{\today}

\begin{abstract}
The magnetic field induced rearrangement of the cycloidal spin structure in ferroelectric mono-domain single crystals of the room-temperature multiferroic BiFeO$_3$ is studied using small-angle neutron scattering (SANS). The cycloid propagation vectors are observed to rotate when magnetic fields applied perpendicular to the rhombohedral (polar) axis exceed a pinning threshold value of $\sim$5\,T. In light of these experimental results, a phenomenological model is proposed that captures the rearrangement of the cycloidal domains, and we revisit the microscopic origin of the magnetoelectric effect. A new coupling between the magnetic anisotropy and the polarization is proposed that explains the recently discovered magnetoelectric polarization to the rhombohedral axis.
\end{abstract}

\maketitle

Owing to the strong cross coupling between magnetism and electric polarization several potentially ground breaking applications of magnetoelectric (ME) multiferroics are suggested such as magnetic read heads, magnetoelectric memory or logic devices \cite{Martin2010,Sando2013,Heron2014}. Among multiferroics BiFeO$_3$ is by far the most studied compound due to its large ferroelectic polarization \cite{Wang2003,Ito2011}, switchable ferroelectric domains \cite{Wang2003,Ito2011,Seongsu2008} and multiferroic phase at room temperature \cite{Roginskaya1966,Fischer1980}, all of which are crucial for applications. Although this material has been studied for half a century, the origin of ME coupling in bulk crystals is still under debate due to both the low-symmetry of Fe sites and Fe-Fe bonds, and the complex magnetic order.

A rhombohedral distortion of the perovskite structure of BiFeO$_3$, which reduces the space group symmetry to $R3c$, generates the large ferroelectric polarization $P$$\approx$40\,$\mu C/cm^2$ along the $\langle$111$\rangle$-type directions below $T_C$=1100\,K \cite{Roginskaya1966,Fischer1980,Seongsu2008,Wang2003,Ito2011}. At $T_N$=640\,K the $S$=5/2 iron spins order into a G-type antiferromagnetic (AF) structure, thus BiFeO$_3$ becomes multiferroic \cite{Roginskaya1966,Fischer1980}. The ferroelectric distortion induces a uniform Dzyaloshinskii-Moriya interaction (DMI), which modifies the AF order with a superposed, long wavelength ($\lambda$=62\,nm) cycloidal modulation \cite{Sosnowska1982,Lebeugle2008}. The propagation vectors ($\textbf{q}$) of the cycloids are aligned with $\langle$1$\overline{1}$0$\rangle$ directions in the planes perpendicular to the polar axis. Due to this ME coupling, which is referred to as the spin-current \cite{Katsura2005} or spin flexoelectric interaction \cite{Sparavigna1994}, the rotation sense of the AF spin cycloid is uniquely defined \cite{Johnson2013}. In the polar phase the lack of inversion symmetry further leads to a second staggered DMI \cite{Ederer2005}, which results in a 1$^\circ$ canting of the AF order perpendicular to the cycloidal plane as visualized in Fig.~1 of Ref.~\onlinecite{Ramazanoglu2011}. This weak-ferromagnetic component enables the observation of the magnetic order and domain populations by small-angle neutron scattering (SANS) \cite{Ramazanoglu2011}.

\begin{figure*}[t]
\includegraphics[width=\textwidth]{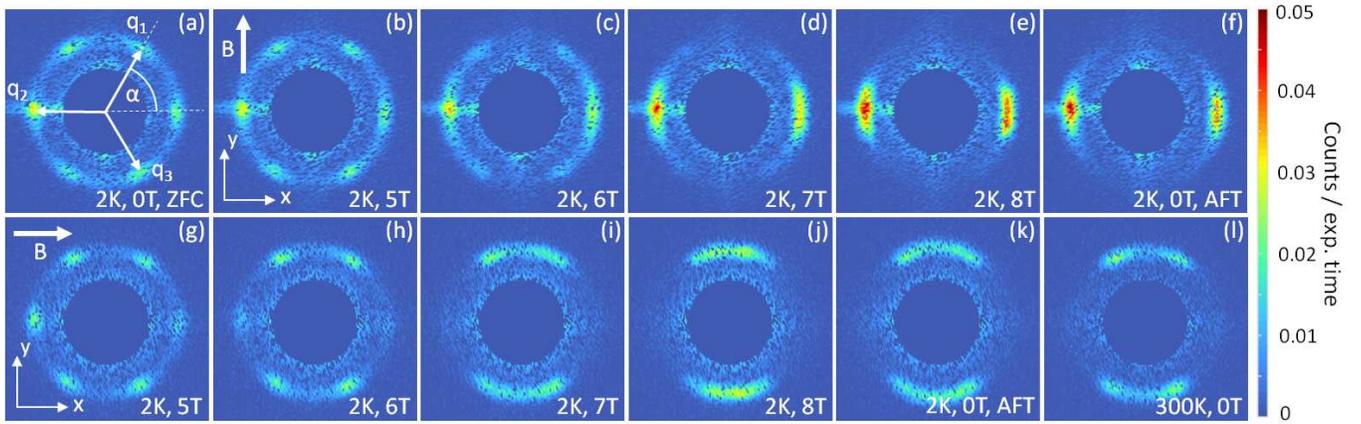}
\caption{(color online) (a)-(e) Magnetic field dependence of the SANS patterns recorded in the $\textbf{z}$ or (111) plane, when a magnetic field was applied along the $H\parallel$$\textbf{y}$ axis ([11$\overline{2}$] direction) in the zero field cooled (ZFC) state at 2\,K. (f) Zero field image after the field treatment (FT). (g)-(j) The field dependence measured for $H\parallel$$\textbf{x}$ ([1$\overline{1}$0] direction) after the sample had been heated above $T_N$ and zero-field cooled. After subsequent removal of the field SANS patterns were recorded at (k) 2\,K and then at (l) 300\,K, respectively.}
\label{Fig1}
\end{figure*}

Recently, advances in the synthesis of single ferroelectric domain single crystals allowed the detailed investigation of the origin of the ME coupling \cite{Seongsu2008,Ito2011}. High-resolution neutron diffraction studies of the crystal structure indicated that a decrease of the ferroelectric polarization below $T_N$ may be attributed to negative magnetostriction \cite{Sanghyun2013}. On the other hand a more recent systematic study of the magnetic field-induced excess polarization, $\Delta$$\textbf{P}$ measured along all the principal crystallographic axes was interpreted in terms of an antisymmetric exchange mechanism in Ref.~\onlinecite{Tokunaga2015}. Following their conventions our Cartesian coordinate system, $\textbf{x}$, $\textbf{y}$, $\textbf{z}$ is respectively fixed to [1$\overline{1}$0], [11$\overline{2}$] and the ferroelectric [111] axes in the pseudocubic notation. It was also reported that at a critical magnetic field the cycloidal structure is unwound and a two sublattice canted AF order is stabilized \cite{Tokunaga2015,Ruette2004}. Across this meta-magnetic transition $\Delta P_z$ changes rapidly, which indicates the cycloid to carry a significant polarization compared with the canted AF state, in agreement with the coupling between the ferroelectric polarization and the uniform DMI. However, $\Delta P_x$ and $\Delta P_y$ are also different in the two phases, which cannot be explained by the usual spin-current model \cite{Katsura2005,Sparavigna1994}. Therefore, more general antisymmetric exchange terms $d$$_{\beta\alpha}$($\mathbf{S}$$_i$$\times$$\mathbf{S}$$_j$)$_\alpha$ were proposed to describe excess polarization both along and perpendicular to the polar axis \cite{Tokunaga2015}.

Theoretical works commonly assume that in BiFeO$_3$ the cycloidal q-vectors are fixed to the $\langle$1$\overline{1}$0$\rangle$-type directions by strong magnetocrystalline anisotropy \cite{Tokunaga2015,Fishman2013}. Therefore, an applied magnetic field can only tune the length of the q-vectors and alter the population of the three $\langle$1$\overline{1}$0$\rangle$-type domains. In addition, these theories fail to capture the zero-field hysteresis observed in either $P_y$ or the infrared absorption spectrum of the spin-wave excitations after the application of high magnetic fields \cite{Nagel2013}.

In this Letter, we report a SANS study of BiFeO$_3$, which shows that moderate magnetic fields of $\mu_0H$$\gtrsim$5\,T can overcome the magnetocrystalline anisotropy and rotate the q-vectors in the plane perpendicular to the polar axis. The changes in the orientation of $\textbf{q}$ and in the population of the domains are irreversible, which explains the aforementioned hysteresis effects. Most importantly, our SANS data motivate the proposal for a new ME mechanism in BiFeO$_3$ giving rise to a polarization perpendicular to the polar axis.

SANS experiments were carried out on a ferroelectric mono-domain BiFeO$_3$ single crystal of 25\,mg grown by a LASER floating-zone technique. The details of the crystal growth reported in Ref.~\cite{Ito2011}. SANS were performed using the D33 instrument of the Institut Laue-Langevin, and the SANS-I instrument of the Paul Scherrer Institut. The incoming neutron wavelength was always set to 8\,{\AA}.

Typical SANS patterns shown in Fig.~\ref{Fig1} represent the sum of detector images, which were measured for an incoming neutron beam nearly parallel to the $\textbf{z}$ axis and as the sample was rotated (rocked) around both the $\textbf{x}$ and the $\textbf{y}$ axes by $\pm$3$^\circ$ in 0.2$^\circ$ steps. After the initial cooling of the sample to 2\,K, when the as-grown crystal had not been exposed to any magnetic fields before, six Bragg-spots were resolved with their q-vectors aligned with the symmetry equivalent $\langle$1$\overline{1}$0$\rangle$ directions (Fig.~\ref{Fig1} a). Each spot pair at $\pm\mathbf{q}$ is due to one of the three cycloidal domains labelled as $q_1$, $q_2$ and $q_3$ (see Fig.~\ref{Fig1}), and their relative intensities reflect the population of the domains, which are nearly equal in the as-grown state. From the peak position in the $|\mathbf{q}|$ dependence of the intensity the periodicity of the cycloid is determined to be 62.2\,nm, which agrees well with the cycloidal wavelength observed previously in BiFeO$_3$ using neutron diffraction and SANS \cite{Sosnowska1982,Lebeugle2008,Ramazanoglu2011}.

Next, magnetic fields were applied along the $\textbf{y}$ direction, which did not change the scattering pattern up to 5\,T (Fig.~\ref{Fig1} b). However, the intensity of the spots in the $q_1$ and $q_3$ domains gradually decreased for $\mu_0H$$\gtrsim$5\,T (Fig.~\ref{Fig1} c-e). In high magnetic fields only the $q_2$ domain remained with $\textbf{q}$ perpendicular to the field. Moreover, the initial domain population was not restored when the field was decreased to zero at 2\,K (Fig.~\ref{Fig1} f) and not even when the sample was warmed up to 300\,K in zero-field (not shown). The initial domain population was reset only after the sample was heated above $T_N$, to $T$=673\,K.

For $\textbf{H}\parallel\textbf{x}$, similarly to $\textbf{H}\parallel\textbf{y}$, no change in the SANS pattern was observed up to 5\,T (Fig.~\ref{Fig1} g). In higher fields the population of the $q_2$ domain with $\textbf{q}$ parallel to the field gradually fell, whereas $\textbf{q}_1$ and $\textbf{q}_3$ rotated toward the $\textbf{y}$ direction perpendicular to the field (Fig.~\ref{Fig1} h-j). When the field was decreased to zero, these q-vectors relaxed slightly back to their initial positions (Fig.~\ref{Fig1} k). This rearrangement was more pronounced after warming the sample up to 300\,K (Fig.~\ref{Fig1} l). On the other hand the $q_2$ domain remained unpopulated even at 300\,K.

\begin{figure}[t]
\includegraphics[width=2.8in]{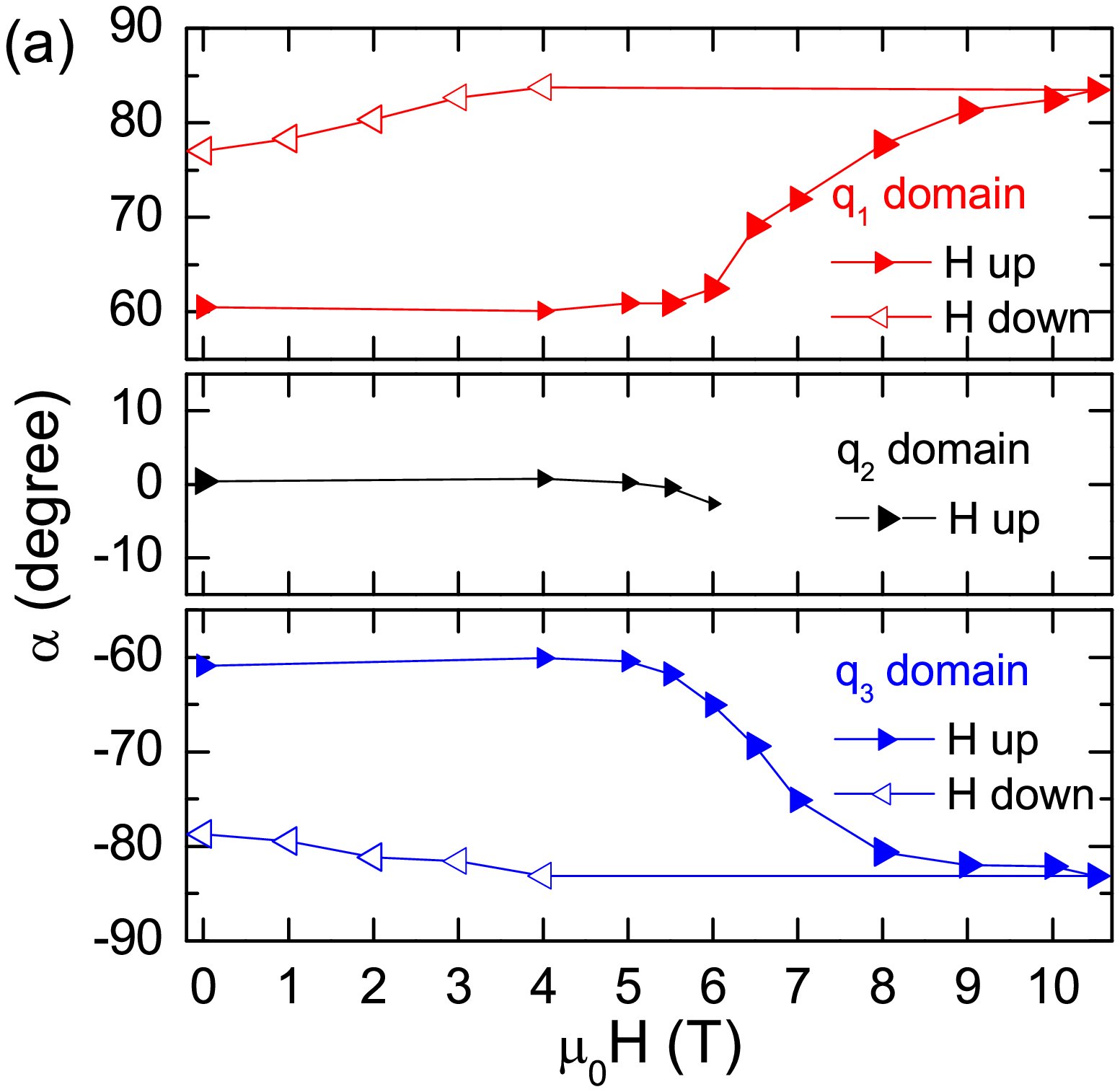}\\
\vspace{1pt}
\includegraphics[width=2.8in]{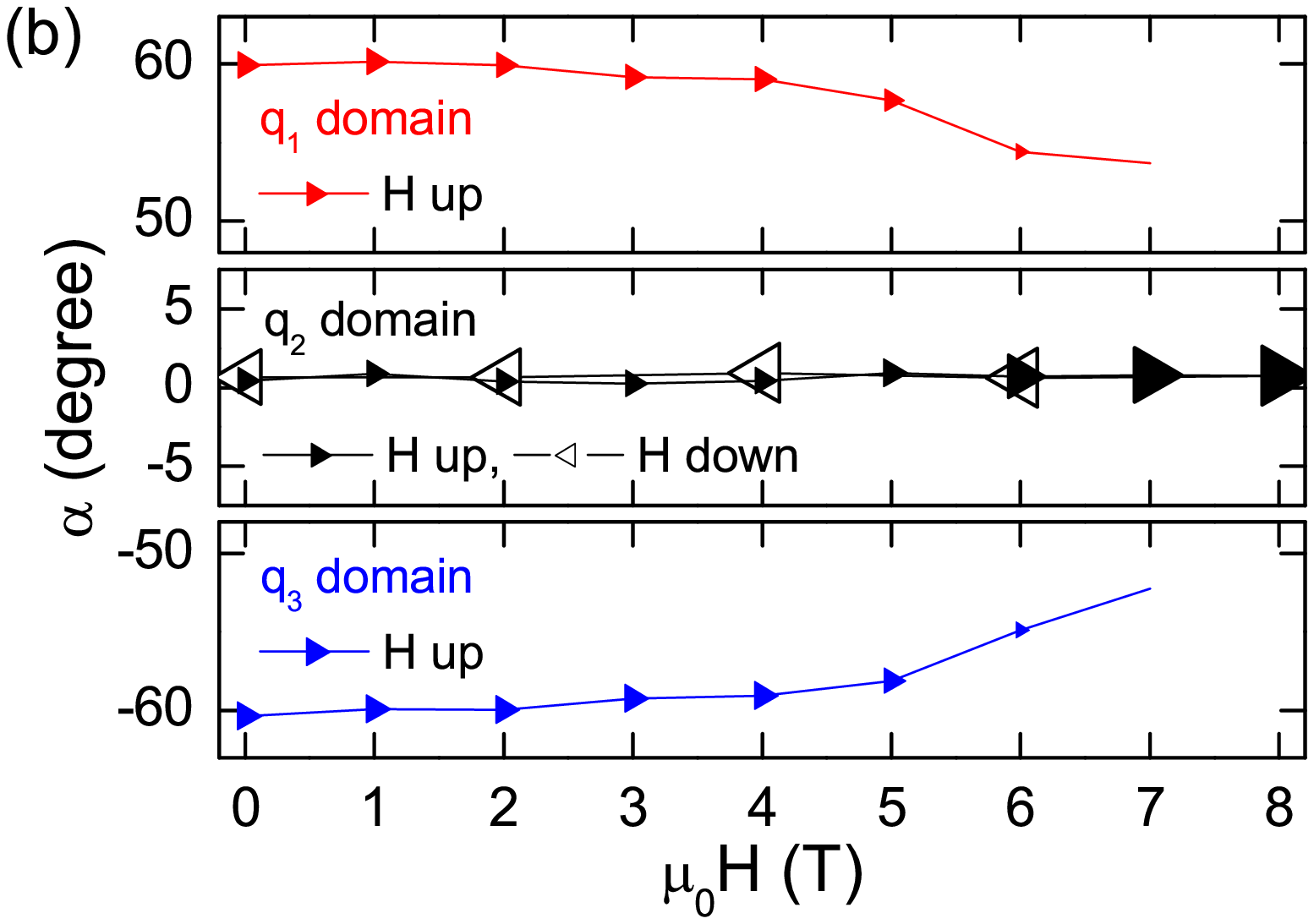}
\caption{(color online) Magnetic field dependence of the azimuthal position of $\textbf{q}$ within the $\textbf{z}$ plane, for (a) $\textbf{H}\parallel\textbf{x}$ and (b) $\textbf{H}\parallel\textbf{y}$, respectively. The azimuthal angles of $\pm$$\textbf{q}$ are averaged according to ($\alpha_q$+$\alpha_{-q}$$-$180$^\circ$)/2 in each domain, with $\alpha$ as defined in Fig.~\ref{Fig1} (a). The symbol sizes are proportional to the scattered intensity of the corresponding peaks.}
\label{Fig2}
\end{figure}

To quantitatively analyse field dependent changes in the Bragg-peak positions, their azimuthal angle $\alpha$ measured from the $\textbf{x}$ axis (Fig.~\ref{Fig1} a) was deduced by fitting the azimuthal angle-dependent total scattered intensity with a sum of Gaussian peaks. The azimuthal angles averaged according to $\alpha=$($\alpha_q$+$\alpha_{-q}$$-$180$^\circ$)/2 for $\pm$$\textbf{q}$ of the same domain are shown in Fig.~\ref{Fig2}. For both field directions, $\textbf{H}\parallel\textbf{x}$ and $\textbf{H}\parallel\textbf{y}$, the angular positions of $\textbf{q}$ only start to change above 5\,T as seen in Figs.~\ref{Fig2} (a) and (b), respectively. When the magnetic field is applied along the $\textbf{x}$ direction, $\textbf{q}_1$ and $\textbf{q}_3$ rotate toward the $\textbf{y}$ axis. Although these domains are stable for $\textbf{H}\parallel\textbf{x}$ up to the highest fields, their q-vectors do not merge even at 10.5\,T. For  $\textbf{H}\parallel\textbf{y}$ $\textbf{q}_1$ and $\textbf{q}_3$ also move closer to the direction perpendicular to the field, i.e. to $\textbf{x}$ axis, before the corresponding domains disappear.

Following former works describing the magnetic field induced reorientation of $\textbf{q}$ in helimagnets \cite{Plumer1981,Grigoriev2006,Bauer2017}, here we introduce a phenomenological model to describe the energy of a cycloidal domain as a function of the orientation of $\textbf{q}$. A related microscopic theory based on numerical energy minimization has been proposed very recently by Fishman \cite{FishmanArXiv}. In BiFeO$_3$ the uniform DMI favours q-vectors lying in the plane perpendicular to the polar axis \cite{Sparavigna1994}. The in-plane anisotropy can be described by an anisotropy energy $\mathcal{H}_{ani}$\,=\,$-K_6\cos(6\alpha)$, dependent on the azimuthal angle $\alpha$ in accord with the $C_{3v}$ point symmetry of the lattice. In small magnetic fields the Zeeman energy of the cycloid can be described by its linear susceptibility, with components parallel and perpendicular to the cycloidal plane denoted as $\chi_{\parallel}$ and $\chi_{\perp}$, respectively. The angular dependent part of the Zeeman energy has the following form: $\mathcal{H}_{H}$\,=\,$\pm\frac{\mu_0}{4}\Delta\chi H^2\cos(2\alpha)$, where $H$ is the strength of the external field, $\Delta\chi$\,=\,$\chi_{\perp}$-$\chi_{\parallel}$ and $\mu_0$ is the vacuum permeability. The plus and minus signs correspond to field directions $\mathbf{H}\parallel$$\textbf{x}$ and $\mathbf{H}\parallel$$\textbf{y}$, respectively. The total energy of the cycloid is $\mathcal{H}$\,=\,$\mathcal{H}_{ani}$+$\mathcal{H}_{H}$ and a single dimensionless parameter $\beta$\,=\,$H\sqrt{\frac{\mu_0\Delta\chi}{4K_6}}$ describes the orientation of $\textbf{q}$. Since the q-vectors align parallel to $\langle$1$\overline{1}$0$\rangle$ directions in the zero field state of BiFeO$_3$ $K_6$$>$0. In finite fields $\textbf{q}$ tends to perpendicular to $\mathbf{H}$, i.e.~$\Delta\chi$$>$0. Indeed $\chi_{\perp}$ which corresponds to a conical distortion of the cycloid is larger than $\chi_{\parallel}$ which describes its anharmonic distortion.

\begin{figure}[t]
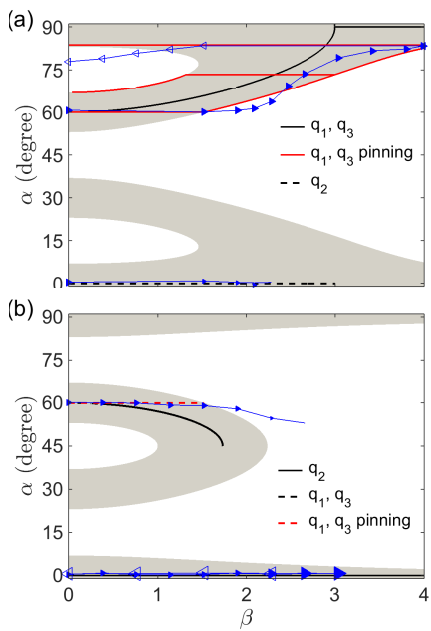

\includegraphics[width=2.4in]{Pin1m10.eps}
\\
\vspace{-21pt}
\includegraphics[width=2.4in]{Pin11m2.eps}
\caption{(color online) Theoretical field dependence of the azimuthal position of $\textbf{q}$ when (a) $\textbf{H}\parallel\textbf{x}$ and (b) $\textbf{H}\parallel\textbf{y}$, respectively. Black solid and dashed lines show the field dependence  of the global and local energy minima, respectively. The grey area shows the region where $\textbf{q}$ is pinned by disorder induced static friction. The red curves show the predicted change in the in-plane orientation of $\textbf{q}$ in the presence of disorder for cases where the highest applied magnetic field is different. The blue symbols represents the measured positions as described in the text.}
\label{Fig4}
\end{figure}

The magnetic field dependence of $\alpha$, as deduced from this model for $\mathbf{H}\parallel$$\textbf{x}$, is shown in Fig.~\ref{Fig4} (a). Due to the symmetry of the model the orientations of $\textbf{q}$ are simply transformed to the 0-90$^\circ$ interval by taking $\alpha_{q_2}$-180$^\circ$ and -$\alpha_{q_3}$ for the $q_2$ and $q_3$ domains, respectively. The solid black line shows the field dependence of the global energy minimum. The q-vectors of the two stable domains $q_1$ and $q_3$ gradually rotate towards the $\pm\textbf{y}$ directions with increasing field, and above a critical field, $\beta_{cx}$\,=\,3, $\textbf{q}$ becomes exactly perpendicular to the field. There is also a local minimum (dashed black line in Fig.~\ref{Fig4} a) which corresponds to the metastable domain $q_2$. The orientation of this q-vector is independent of the field and above $\beta$\,$>$\,$\beta_{cx}$ this domain becomes unstable and disappears.

This model qualitatively captures the observed rotation of $\textbf{q}$ and the changes in the domain population, though it cannot explain the pinning of $\textbf{q}$. The effect of static friction due to impurities is therefore included in the model by assuming that $\textbf{q}$ is not rotated until a threshold torque value, $\tau$ is reached. In Fig.~\ref{Fig4} the region where $\tau>$$\frac{\partial\mathcal{H}}{\partial\alpha}$, thus where the q-vectors are pinned, is shaded in grey. Therefore, when a magnetic field $\mathbf{H}\parallel$$\textbf{x}$ is applied to the as-grown state (red curves in Fig.~\ref{Fig4} a), the orientation of $\textbf{q}$ at 60$^\circ$ is not expected to change until the static friction is overcome. Only then $\textbf{q}$ slips and $\alpha$ increases following the border of white and grey regions corresponding to the threshold torque value. We also note that due to pinning $\alpha$ is not expected to ever be completely perpendicular to the field, which is consistent with our observations. Upon decreasing the field, the high-field value of $\alpha$ is maintained until the small white pocket is reached at lower fields, and where the orientation of $\textbf{q}$ relaxes along the border. In reality the cycloidal state cannot rigidly rotate over infinitely large regions, but instead the sample volume becomes divided to smaller regions. The different pinning torque in these regions can explain the broadening of the spots in field (see Fig.~\ref{Fig1}) and the slight difference between the theoretical curves in Fig.~\ref{Fig4} a and the mean azimuthal position of the q-vectors determined from the experiments.

For the orthogonal field direction $\mathbf{H}\parallel$$\textbf{y}$, as shown in Fig.~\ref{Fig4} (b), the $q_1$ and $q_3$ domains are metastable and their q-vectors are expected to rotate with increasing field until they reach +45$^\circ$ and -45$^\circ$, respectively.
Due to pinning these q-vectors do not even reach these limiting angles up to the critical field, $\beta_{cy}$\,=\,$\sqrt{3}$ where these domains become unstable and disappear. The q-vectors of the $q_2$ domain maintain their zero field alignment, and remain stable up to the highest fields.

Since the reconstruction of the magnetic domain structure can affect the measured bulk electric polarization, in the light of the present SANS results we reconsider the analysis of recent high-field magnetization and magnetic field induced electric polarization data reported in Ref.~\onlinecite{Tokunaga2015}. Experimentally, it was found that the critical field necessary to transform the low field cycloidal phase into a canted antiferromagnetic order is isotropic in the plane perpendicular to the polar axis. This lies in stark contrast to predictions based on microscopic theory \cite{Fishman2013,Tokunaga2015} where the q-vectors are expected to be tightly fixed to the $\langle$1$\overline{1}$0$\rangle$-type directions by the strong in-plane anisotropy. Our experiment first demonstrates that this widely accepted theoretical assumption is not valid and for $\mu_0H$$\gtrsim$7\,T only domains with $\textbf{q}$ nearly perpendicular to the field survive, irrespective of the in-plane direction of the field. Furthermore, the in-plane magnetization and the y component of the magnetic field induced polarization, $\Delta P_y$ show hysteresis below $\lesssim$6\,T \cite{Tokunaga2015} whose origin has not been clarified. Our SANS experiments give compelling evidence that fields on this scale irreversibly depopulate the unfavoured cycloidal domain(s), thus, the hysteresis of $\Delta P_y$ can be attributed to the polarization difference between the as-grown and the field treated states with different domain populations. The model proposed in Ref.~\onlinecite{Tokunaga2015} can reproduce magnetically induced polarizations in the $\textbf{z}$ plane, perpendicular to $\textbf{q}$. However, our experiments evidence a rotation of $\textbf{q}$ which would imply the cancellation of $\Delta P_y$ in their model when the external field is applied along the $\textbf{x}$ direction.

To resolve this contradiction we propose a Landau theory where the coupling between the in-plane polarization, $\textbf{p}$ and the antiferromagnetic vector, $\textbf{l}$ is described by the free energy density term $\Phi_{ME}$=$\gamma$($p_xl_xl_z+p_yl_yl_z$)+$\delta$($p_xl_xl_y+p_y\frac{l_x^2-l_y^2}{2}$) in the lowest order. We note that a similar expression was derived in Ref.~\onlinecite{Rovillain2010,deSousa2013} to describe the electric field induced shift of the spin-wave energies, but it contained electric field components instead of the ferroelectric polarization. The first term with coefficient $\gamma$ cancels for a cycloid since it only gives an oscillating polarization. However, the $\delta$ term describes an in-plane polarization induced by the antiferromagnetic cycloid with non-trivial dependence on the azimuthal angle: [$p_x$, $p_y$] $\propto$ [$\sin(2\alpha)$, $\cos(2\alpha)$]. In the as-grown state with equally populated domains the in-plane polarization components induced by the different domains cancel out. After exposing the material to $\mathbf{H}\parallel$$\textbf{y}$ a finite polarization arises only in the $\textbf{y}$ direction, whereas $\mathbf{H}\parallel$$\textbf{x}$ induces polarization along the -$\textbf{y}$ direction in agreement with the polarization measurements. When the q-vectors are perpendicular to the field $\Delta P_y$ is expected to have opposite sign but the same magnitude for $\mathbf{H}\parallel$$\textbf{x}$ and $\mathbf{H}\parallel$$\textbf{y}$, while experimentally $\Delta P_Y$=200\,$\mu$C/m$^2$ and -450\,$\mu$C/m$^2$ is found, respectively. The difference in the absolute values can be explained by higher order anisotropy terms or by the fact that the q-vectors are not completely perpendicular to the field for $\mathbf{H}\parallel$$\textbf{x}$. Besides $\Delta P_Y$, a transverse polarization, which vanishes for $\mathbf{H}\parallel$$\textbf{x}$ and $\mathbf{H}\parallel$$\textbf{y}$ but which is finite for intermediate orientations of the field is expected according to our model. To the best of our knowledge such an experiment has never been performed though it may unveil further details of the ME coupling in BiFeO$_3$.

Microscopically, the ME free energy term $\Phi_{ME}$ introduced above can be associated with anisotropic magnetostriction or an on-site metal-ligand hybridization \cite{deSousa2013}. We expect that this mechanism dominates over the antisymmetric $d$$_{\beta\alpha}$($\mathbf{S}$$_i$$\times$$\mathbf{S}$$_j$)$_\alpha$-type terms proposed in Refs.~\onlinecite{Tokunaga2015,Miyahara2016}. The former involves third order terms of $l_\alpha$ and $p_\alpha$, whereas the latter corresponds to fourth-order terms also involving spatial derivatives of $l_\alpha$.

In summary, we studied the magnetic field-driven rearrangement of cycloidal domains in BiFeO$_3$ using SANS. For in-plane fields we found that above $\gtrsim$7\,T only the domains favoured by the external field survive. Moreover, if the alignment of the q-vectors in zero field is not perpendicular to the magnetic field, they tend to rotate orthogonal to the field above $\gtrsim$5\,T when the static friction due to disorder induced pinning is overcome. Based on the rotation of $\textbf{q}$ the origin of the spin-polarization coupling is reconsidered and a new ME term is introduced, this amounting to an important step forward for a complete theory of the ME effect in BiFeO$_3$.

\begin{acknowledgments}
We grateful to R.~S.~Fishman, T.~R\~o\~om, U.~Nagel, D.~Szaller for fruitful discussions. This work was supported by Hungarian Research Funds OTKA K 108918, OTKA PD 111756, National Research, Development and Innovation Office – NKFIH, ANN 122879, Bolyai 00565/14/11, the Swiss National Science Foundation (SNF) Sinergia network 'NanoSkyrmionics', and the SNF project grant 153451, the Deutsche
Forschungsgemeinschaft (DFG) via the Transregional Research Collaboration TRR 80: From Electronic Correlations to Functionality (Augsburg - Munich - Stuttgart). This work is based on neutron experiments performed at the Institut Laue-Langevin (ILL), Grenoble, France and the Paul Scherrer Institut, Villigen, Switzerland. A portion of this research used resources at the High Flux Isotope, a DOE Office of Science User Facility operated by the Oak Ridge National Laboratory.
\end{acknowledgments}

\end{document}